\def\articleTitle{Gaussian distributional structural equation models: A framework for modeling latent heteroscedasticity}

    \documentclass[letterpaper,12pt]{article}
  
    \usepackage{abstract} % Allows abstract customization
    \usepackage{titlesec} % Allows customization of titles

    \usepackage{authblk} % For multiple authors

	\usepackage{datetime} % allows for including today's date
    \newdateformat{usvardate}{
  	\monthname[\THEMONTH] \ordinal{DAY}, \THEYEAR}

  	\usepackage[bottom]{footmisc} % Makes footnotes stick to bottom of the page
    
    \usepackage{fancyhdr}
    \fancyheadoffset{0cm}
    \providecommand{\keywords}[1]{\textbf{\textit{Keywords---}} #1}

\newcommand\wordcount{%
  \immediate\write18{texcount -utf8 -merge -sum -incbib -dir -sub=none -brief \jobname.tex | cut -d : -f 1 > 'count.txt'}%
  \input{count.txt}\ignorespaces words%
}

    \usepackage{amssymb}
    \usepackage{amsthm}
    \usepackage{newtxmath}
    \usepackage{bm}
    \usepackage{upgreek}

    \usepackage{microtype} % Slightly tweak font spacing for aesthetics
    \usepackage[utf8]{inputenc}
    \usepackage{newtxtext} % Makes default font Adobe Times New Roman
  
    \usepackage{setspace} % See \doublespacing command at the top of content.tex
    \usepackage{lineno,xcolor} 	% See \linenumbers at the top of content.tex

	\usepackage[top=1.5in, bottom=1.5in, left=1in, right=1in]{geometry}

	\usepackage[colorinlistoftodos]{todonotes} % allows margin comments
	\usepackage{soul} % allows for highlighting
    
    \usepackage{graphicx} % More advanced figure inclusion
    \usepackage{float} % For specifying table/figure locations, i.e. [ht!]
    \usepackage{tikz}
    \usepackage{tikz-cd}
    \usepackage{comment}
    \usetikzlibrary{positioning}
    \usetikzlibrary{bayesnet}
    \usetikzlibrary{arrows}
        
    \usepackage{printlen}

    \usepackage{longtable} % For long tables that span multiple pages
    \usepackage{tabularx} % Allows advanced table features
    \newcolumntype{L}[1]{>{\raggedright\arraybackslash}p{#1}}
    \newcolumntype{C}[1]{>{\centering\arraybackslash}p{#1}}
    \newcolumntype{R}[1]{>{\raggedleft\arraybackslash}p{#1}}
    \usepackage{relsize} % Allows precise adjustment of font size,
    \usepackage{tabularray} % See: https://www.latex-tables.com/ressources/tabularray.html

    \usepackage{hyperref} % For hyperlinks in the PDF
    \usepackage{csquotes}
    \usepackage[style=apa6,backend=biber]{biblatex}
\begin{document}

% TITLE SECTION

% Comment out one of the two lines below to include or exclude authors in article title. You may want to exclude them in case of a double-blind peer review submission, since authors appear on the cover page, and reviewers should not see the authors' names in the manuscript draft.

%--------------------------------------------------------%
%	TITLE SECTION with authors
%--------------------------------------------------------%

% Article title
  \title{\vspace{-15mm}\fontsize{21pt}{10pt}\selectfont\textbf{\articleTitle}}

% Authors and Affiliations
  \author{Luna Fazio$^{1*}$, Paul-Christian Bürkner$^{1}$}
  \renewcommand\Authands{ and }

\date{\small
  $^1$ Department of Statistics, TU Dortmund University, Germany\break
  $^*$ Corresponding author; Email: \texttt{bmfaziol@gmail.com}
}

\maketitle % Insert title

%--------------------------------------------------------%
%	CONTENT
%--------------------------------------------------------%

%--------------------------------------------------------%
%	ABSTRACT
%--------------------------------------------------------%

\begin{abstract}
\noindent Accounting for the complexity of psychological theories requires methods that can predict not only changes in the means of latent variables -- such as personality factors, creativity, or intelligence -- but also changes in their variances.

Structural equation modeling (SEM) is the framework of choice for analyzing complex relationships among latent variables, but the modeling of latent variances as a function of other latent variables is a task that current methods only support to a limited extent.

In this paper, we develop a Bayesian framework for Gaussian distributional SEM which broadens the scope of feasible models for latent heteroscedasticity. We use statistical simulation to validate our framework across four distinct model structures, in which we demonstrate that reliable statistical inferences can be achieved and that computation can be performed with sufficient efficiency for practical everyday use. We illustrate our framework's applicability in a real-world case study that addresses a substantive hypothesis from personality psychology.
\end{abstract}

% Insert keywords here
\setlength\parindent{0in} \keywords{structural equation modeling, distributional regression, Bayesian inference, heteroscedasticity, measurement invariance}

%--------------------------------------------------------%
%	BODY TEXT
%--------------------------------------------------------%
% Start double spacing here if you want
%\doublespacing

\section{Introduction}
\label{sec:intro}

Structural equation modeling (SEM) is a widely-used statistical framework that can be regarded as an extension of regression models: it allows modeling multiple dependent variables simultaneously, including relationships among them, as well as the introduction of measurement error and unobserved (latent) variables (for a comprehensive introduction see \cite{bollenStructuralEquationsLatent1989,klinePrinciplesPracticeStructural2016}).
As with regression, the classic formulation of SEM presents an idealized setting where, among other simplifications, it is assumed that the model's parameters (intercepts, coefficients and (co-)variances) all take on values that are constant across people, conditions, etc. Such an assumption often does not hold in practice and this has motivated a rich literature on methods for handling non-invariant parameters (see below). Extending this line of research, we develop a Bayesian framework for Gaussian distributional SEMs, which, compared to past approaches, supports more flexible models of latent heteroscedasticity when dependencies on other latent variables are involved. We demonstrate our framework's statistical validity and usefulness through simulation studies on four distinct structural models and a real-world case study applied to a research question from personality psychology.

\subsection{Related work}
\label{sec:relatedwork}

The problem of invariance has received attention since the early days of factor analysis, initially focusing on invariance of the covariance matrix of observed data across selected subgroups of some larger population \parencite{thomsonInfluenceMultivariateSelection1939}. The introduction of the multiple-group model methodology in \textcite{joreskogSimultaneousFactorAnalysis1971} marked the shift in focus to the invariance of model parameters that prevails today. It was followed by the development of \emph{moderated factor analysis} (MFA), which enabled modeling parameter values via known functions of observed variables (moderators), including continuous ones; this meant that evaluation of invariance stopped being limited to comparisons over discrete groups \parencite{bauerPsychometricApproachesDeveloping2009}. A further extension, \emph{local structural equation modeling} (LSEM), fits the model multiple times over the moderators' range in combination with an observation weighting scheme to produce a nonparametric estimate of the moderation functions, thereby avoiding the assumption of a known functional form \parencite{hildebrandtExploringFactorModel2016}. There are additional approaches which are particularly suited for assessment of non-invariance under specific assumptions of magnitude or structure \parencite[for an overview see][]{leitgobMeasurementInvarianceSocial2023}, but we do not discuss them here as they are less related to our proposed framework.

The above-mentioned techniques already provide a great deal of flexibility for modeling varying parameters within the SEM framework, but they all share the requirement that the moderator be an observed variable. A set of related approaches known as \emph{heteroscedastic factor models} use MFA-like regressions on residual item variance and factor loading parameters together with skewed latent variable distributions \parencite{molenaarTestingModellingNonnormality2010,molenaarModelingAbilityDifferentiation2011}. Another approach introduced in \textcite{molenaarHeteroscedasticLatentTrait2015} is to use latent skewed distributions to allow the model to account for the effects of continuous latent moderators on the latent trait of interest. However, it achieves so by effectively marginalizing over the moderator and hence is not applicable when one wishes to include a measurement model for the latent moderator. In his discussion, \citeauthor{molenaarHeteroscedasticLatentTrait2015} mentions this limitation and notes that models with explicit latent moderators would constitute a useful addition to the literature, citing methods for investigating latent heteroscedasticity \parencite{molenaarDetectingSpecificGenotype2012} and latent variable interactions \parencite{kleinMaximumLikelihoodEstimation2000} as examples.

Indeed, one can already find some developments towards the use of latent predictors for latent variances. For instance, the works of \textcite{nestlerModellingInterindividualDifferences2020} and \textcite{martinReliabilityFactorModeling2022}, motivated from the perspective of measurement reliability, provide techniques for modeling the variance of measurement errors (which can be conceived of as a special type of latent variable) as dependent on other latent variables. The original formulation of MFA \parencite{bauerPsychometricApproachesDeveloping2009} also received an extended treatment in \textcite{bauerMoreGeneralModel2017}, where it is emphasized that the method can be used to assess measurement invariance and differential item functioning, including the case of both observed and latent moderators of item-level residual variances. Moving beyond heteroscedastic errors, modeling of the residual variance of a structural latent variable has also been demonstrated under a frequentist framework in a simple latent regression setting (i.e. one exogenous latent variable predicting one endogenous latent variable; \cite{dekortStudyingStrengthPrediction2017}).

One key challenge in maximum likelihood estimation of SEM is that latent variables can be regarded as \emph{incidental} \parencite{neymanConsistentEstimatesBased1948} or \emph{nuisance} \parencite{basuEliminationNuisanceParameters1977} parameters, which means that they must be marginalized out of the likelihood in order for consistent estimates to be obtainable. When latent heteroscedasticity is introduced, a closed-form expression of the marginal likelihood will generally not be available and numerical integration has to be performed at each step of the maximization procedure (e.g. \cite{hessenHeteroscedasticOnefactorModels2009}). Such an approach is sometimes called Marginal Maximum Likelihood (MML) and corresponds to an application of the more general expectation-maximization (EM) algorithm \parencite{bockMarginalMaximumLikelihood1981}. As the quadrature methods that are commonly used to approximate the integral do not scale well with dimension (which in turn grows with the number of latent variables), \textcite{dekortStudyingStrengthPrediction2017} have suggested that Bayesian procedures could provide a viable alternative to MML for estimation of larger models with latent heteroscedasticity. To our knowledge, a systematic assessment of such an approach has not yet been conducted.

\subsection{Our contributions}

We develop and validate a Bayesian approach to support latent moderators of latent variances, which works by including latent variables as parameters to sample from instead of marginalizing over them. Such an approach has been termed \emph{conditional likelihood} in the latent variable literature (e.g. \cite{merkleBayesianComparisonLatent2019}) and it was favored in earlier methods for obtaining full posterior distributions in Bayesian SEM as it enabled the use of Gibbs sampling \parencite{leeStructuralEquationModeling2007}. With the development of algorithms such as Hamiltonian Monte Carlo (HMC; \cite{nealMCMCUsingHamiltonian2012,betancourtConceptualIntroductionHamiltonian2018}), it was no longer necessary to use conditional distributions that could be sampled from and contemporary Bayesian SEM software has moved to use marginal likelihoods due to the increased sampling efficiency gained by not having the latent variables as additional parameters \parencite{merkleEfficientBayesianStructural2021}. As mentioned above, however, latent moderators cannot be handled in full generality when using marginal likelihoods, which is why we adopt a conditional likelihood approach in this paper.

We implement our framework in the probabilistic programming language Stan, which provides an expressive syntax and powerful algorithms to specify and fit open-ended Bayesian models \parencite{standevelopmentteamStanModelingLanguage2023}. To avoid users having to interact with Stan directly, we extended the \texttt{R} package \texttt{brms}, designed to simplify the process of fitting Bayesian regression models in Stan  while still providing access to advanced regression techniques that can be combined in a modular fashion \parencite{rcoreteamLanguageEnvironmentStatistical2023,burknerBrmsPackageBayesian2017}. We realize latent variable models with moderators in \texttt{brms} by utilizing its functionality for model-based imputation and distributional regression models, models predicting distributional parameters beyond the mean, for example, also variances or standard deviations (see \cite[Chapter~10]{fahrmeirRegressionModelsMethods2021} and \cite{burknerAdvancedBayesianMultilevel2018,burknerBayesianItemResponse2021}). By representing latent variables as missing observations and placing them as predictors of distributional parameters, we obtain an MFA-like procedure that admits latent moderators with more flexibility than methods based on marginal likelihoods.

In the remainder of this article we describe and evaluate our conditional likelihood approach for continuous latent moderators on both latent means and variances. In Section \ref{sec:model}, we formally introduce the model and establish the corresponding notation. In Section \ref{sec:simstudy}, we present are large-scale simulation study to evaluate our approach, with results showing good convergence and parameter recovery in all investigated models. We demonstrate an application to a substantive hypothesis from personality psychology in Section \ref{sec:casestudy}. Finally, in Section \ref{sec:discussion} we discuss limitations and future directions.

\section{Model description}
\label{sec:model}

Below, we formally introduce the developed SEM framework.
Going forward, we will make an important simplifying assumption: all the variables in the model are conditionally normally distributed. This is not an inherent limitation of the approach we present, as it allows the specification of any continuous distribution for the latent variables, with moderation on other parameters beyond the mean. However, we find that this simplified setting already involves enough complexity for a rich discussion and practical relevance, so we omit a more general treatment in order to keep a reasonable scope for this paper.
Additionally, we will omit structural manifest variables and fixed covariates from the following exposition as their inclusion is straightforward and our interest here is to discuss latent-to-latent regressions.

\subsection{Model likelihood}

We begin by describing the general structure of the model. Let $I$ be some set indexing individual observations over the relevant units of analysis (e.g., institutions, individuals, time points, etc.). For each $i\in I$, we have a vector $\boldsymbol{\zeta}_i = (\zeta_{1i}, \dots, \zeta_{li}, \dots, \zeta_{Li})$ of latent variables and for each $\zeta_{li}$, the vector $\mathbf{y}_{li} = (y_{l1i}, \dots, y_{lmi}, \dots, y_{LMi})$ holds the corresponding manifest indicator variables. Here, $M$ denotes the number of manifest variables of the $l$th factor, with $M$ being allowed to vary over $l$. Then, the distribution of the variables can be written as
\begin{align}
\begin{split}
\zeta_{li}\mid\boldsymbol{\zeta}_i &\sim \text{Normal}(\mu_{li}, \sigma_{li}) \\
y_{lmi}\mid\zeta_{li} &\sim \text{Normal}(\nu_{lm}+\lambda_{lm}\zeta_{li}, \tau_{lm}),
\end{split}
\label{eq:measurementmodel}
\end{align}
where $\nu_{lm}$ is the intercept of the manifest variable and $\lambda_{lm}$ is its factor loading. Both the mean $\mu_{li}$ and standard deviation $\sigma_{li}$ of each latent variable can depend on other latent variables. We consider dependencies of the form given by a generalized additive predictor
\begin{align}
\eta_{\theta_{l}i} =
    \sum_{k=0}^{K_{\theta_l}} \beta_{k{\theta_l}} f_{k{\theta_l}}(\boldsymbol{\zeta}_i),
\end{align}
where $\theta$ stands for the likelihood parameter of interest ($\mu$ or $\sigma$) and $\beta_{k{\theta_l}}$ are the coefficients for each continuous (possibly non-linear) transformation $f_{k{\theta_l}}$ of the latent variables. For clarity, we point out that if one wishes to include an intercept in the model, this can be done by setting $f_{0{\theta_l}} = 1$ and incorporating fixed covariates more generally is a matter of putting their values as a constant part in the $f_{k{\theta_l}}$. With this notation, each parameter is related to its predictor via the appropriate link function:
\begin{align}
\begin{split}
\mu_{li} &= \text{identity}(\eta_{{\mu_l}i}) = \eta_{{\mu_l}i},\\
\sigma_{li} &= \exp(\eta_{{\sigma_l}i}).
\end{split}
\end{align}
Let us set $\boldsymbol{\theta}_{\boldsymbol{\zeta}} = (\beta_{1{\mu_1}}, \dots, \beta_{K{\mu_L}},\beta_{1{\sigma_1}}, \dots, \beta_{K{\sigma_L}})$ to denote the vector of structural parameters and $\boldsymbol{\theta}_{\mathbf{y}} = (\nu_{11},\dots,\nu_{LM},\lambda_{11},\dots,\lambda_{LM},\tau_{11},\dots,\tau_{LM})$ to denote the vector of measurement model parameters. Then, the full likelihood can be written as
\begin{align}
\label{eq:likelihood1}
p(\mathbf{y}, \boldsymbol{\zeta} \mid  \boldsymbol{\theta_{\boldsymbol{\zeta}}}, \boldsymbol{\theta_{\mathbf{y}}}).
\end{align}
Following terminology from the latent variable model literature (e.g. \cite{merkleBayesianComparisonLatent2019}), one could obtain the \emph{marginal likelihood} by integrating out the latent variables:
\begin{align}
\label{eq:likelihood2}
p(\mathbf{y} \mid \boldsymbol{\theta_{\boldsymbol{\zeta}}}, \boldsymbol{\theta_{\mathbf{y}}}) &=
\int p(\mathbf{y}, \boldsymbol{\zeta} \mid \boldsymbol{\theta_{\boldsymbol{\zeta}}}, \boldsymbol{\theta_{\mathbf{y}}}) d\boldsymbol{\zeta}.
\end{align}
The use of marginal likelihoods is a necessity in frequentist settings as latent variables play the role of incidental parameters, which results in inconsistent estimates if they are included in the estimation process (\cite{neymanConsistentEstimatesBased1948}; also see discussion at the end of \cite{hessenHeteroscedasticOnefactorModels2009}). On the other hand, there are no formal impediments for performing Bayesian inference while including latent variables as part of the model parameters. The form of the likelihood in which latent variables are explicitly included is called the \emph{conditional likelihood}, as one can decompose Eq.~\ref{eq:likelihood1} into a likelihood for the indicator variables conditioned on the latent variables, and a likelihood for the latent variables themselves:
\begin{align}
\label{eq:likelihood3}
p(\mathbf{y}, \boldsymbol{\zeta} \mid  \boldsymbol{\theta_{\boldsymbol{\zeta}}}, \boldsymbol{\theta_{\mathbf{y}}}) &= p(\mathbf{y}\mid \boldsymbol{\zeta},\boldsymbol{\theta_\mathbf{y}}) \,
p(\boldsymbol{\zeta}\mid \boldsymbol{\theta_{\boldsymbol{\zeta}}}).
\end{align}
As discussed in Section \ref{sec:intro}, we use conditional likelihoods in this paper because marginalization would not produce a closed-form expression in the presence of latent predictors for latent variances.

Because we use the Bayesian framework for inference (see the \hyperref[sec:estimation]{Estimation} section for more information), a complete specification must also include priors for the parameters. We can write the resulting joint posterior as
\begin{align}
\label{eq:fullpost}
\begin{split}
p(\boldsymbol{\zeta}, \boldsymbol{\theta_{\boldsymbol{\zeta}}}, \boldsymbol{\theta_{\mathbf{y}}}\mid \mathbf{y}) &\propto   
p(\mathbf{y}\mid \boldsymbol{\zeta},\boldsymbol{\theta_\mathbf{y}}) \,
p(\boldsymbol{\zeta}\mid \boldsymbol{\theta_{\boldsymbol{\zeta}}}) \,
p(\boldsymbol{\theta_\mathbf{y}}) \,
p(\boldsymbol{\theta_{\boldsymbol{\zeta}}}) \\
&= \left[\prod_{l=1}^L p(\mathbf{y}_l \mid \zeta_l, \boldsymbol{\theta_\mathbf{y}}) \,
p(\zeta_l \mid \text{PA}(\zeta_l), \boldsymbol{\theta_{\boldsymbol{\zeta}}})\right] \,
p(\boldsymbol{\theta_\mathbf{y}}) \,
p(\boldsymbol{\theta_{\boldsymbol{\zeta}}}),
\end{split}
\end{align}
where $\text{PA}(\zeta_l) \subseteq \boldsymbol{\zeta}_{-l}$ denotes the \emph{parents} of latent variable $\zeta_l$ among the set of all other latent variables $\boldsymbol{\zeta}_{-l}$, that is, all latent variables that contribute to the additive predictors $\boldsymbol{\mu}_{l}$ or $\boldsymbol{\sigma}_{l}$.

\subsection{Identification}

The model as given above is underidentified. Unless otherwise noted, identification for models in this paper is obtained by setting the expectation of all latent variables to 0, which identifies their mean, and the loading factor of one item to 1, which identifies their scale (see \cite[p. 238]{bollenStructuralEquationsLatent1989} for an introduction to identification in SEM).
Our model additionally introduces coefficients for the latent variance linear predictor, so it is valid to ask whether these parameters are identified too. Below, we provide a formal argument demonstrating that a link from observed data to parameter values can be drawn without the need for any new constraints, thus showing identification.

To start with, consider the simplified scenario where we assume the latent values to be observed directly. Let $\zeta_0$ be the variable whose variance we are interested in predicting based on the values of ${\zeta_1, \dots, \zeta_K}$ so that
\begin{align}
\label{ref:latentvariance}
\text{Var} \, \zeta_0 = \exp\left(\Sigma^K_{k=1} \beta_k\zeta_k\right),
\end{align}
which means that the coefficients are related to the ratio of variances given a unit increase in one of the predictors. Without loss of generality, consider increasing $\zeta_1$ by one. Then, we find
\begin{align}
\label{eq:latentratio}
\frac{\text{Var} \left[ \zeta_0 \mid (\zeta_1 + 1), \zeta_{\{2,\dots,K\}} \right]}{\text{Var} \left[ \zeta_0 \mid \zeta_1, \zeta_{\{2,\dots,K\}} \right]} =
\left(\frac{\exp(\beta_1(\zeta_1 + 1) + \Sigma^K_{k=2} \beta_k\zeta_k)}{\exp(\beta_1\zeta_1 + \Sigma^K_{k=2} \beta_k\zeta_k)}\right)
= \exp(\beta_1),
\end{align}
thus showing identification of the coefficients ${\beta_1, \dots, \beta_K}$ if there is variation in the corresponding latent variables. Estimation of the coefficients in such a model is a well-studied topic (e.g. \cite{harveyEstimatingRegressionModels1976}) and can be regarded as a particular case of the more general \emph{distributional regression} framework (e.g., see Chapter 10 of \cite{fahrmeirRegressionModelsMethods2021}).

Coming back to our non-simplified model, we do not actually observe the latent variables but rather noisy measurements as defined in Eq. \ref{eq:measurementmodel}, and we want to show whether it's possible to infer changes in the latent variance from those available observations. For this purpose, a more helpful way of writing the measurement model is
\begin{align}
\label{eq:measurementmodel2}
y_{lm} = \lambda_{lm}\zeta_l + \varepsilon_{lm},\quad\varepsilon_{lm}\sim \text{Normal}(0,\tau_{lm}),
\end{align}
which can be combined with Eq. \ref{ref:latentvariance} to untangle the latent variance from the error variance. Let us examine the observed variance of some measurement $y_{0m}$ of $\zeta_0$, conditional on observed measurements $y_{\{1,\dots,K\}m}$ for the latent predictors. For simplicity, and without loss of generality, we take a single measurement per latent variable so the $m$ subscript is dropped. This gives us
\begin{align}
\text{Var} \left[ y_{0}\mid y_{\{1,\dots,K\}},\varepsilon_{\{1,\dots,K\}} \right] &=
    \text{Var}\left[\lambda_{0}\zeta_0 + \varepsilon_{0}\right] \nonumber \\
\label{eq:latentvariance2}
&= \lambda^2_{0}\text{Var}\zeta_0 + \tau_{0}^2\\
&= \lambda^2_{0}\exp(\Sigma^K_{k=1} \beta_k \zeta_k) + \tau_{0}^2 \nonumber\\
\label{eq:latentvariance3}
&= \lambda^2_{0}\exp(\Sigma^K_{k=1} \beta_k(y_k-\varepsilon_k)/\lambda_k) +
    \tau_{0}^2.
\end{align}
The last expression still contains unobserved variables in the form of the error terms $\varepsilon_{\{1,\dots,K\}}$, but we can apply the law of iterated expectations to deal with them. We use Eqs. \ref{eq:latentvariance2} and \ref{eq:latentvariance3} to work in terms of the latent variance we are interested in:
\begin{align}
\text{Var} \left[ \zeta_0\mid y_{\{1,\dots,K\}} \right] &=
\mathbb{E}_{\varepsilon} \left[
    \text{Var} \left[ \zeta_0\mid y_{\{1,\dots,K\}},\varepsilon_{\{1,\dots,K\}} \right]
\right] \nonumber\\
&= \mathbb{E}_{\varepsilon}\left[
    \exp(\Sigma^K_{k=1} \beta_k(y_k-\varepsilon_k)/\lambda_k)    
\right] \nonumber\\
&= \exp(\Sigma^K_{k=1} \beta_k y_k/\lambda_k)    
\mathbb{E}_{\varepsilon}\left[
\exp(-\Sigma^K_{k=1} \beta_k \varepsilon_k/\lambda_k)    
\right] \nonumber\\
\label{eq:latentvariance4}
&= \exp(\Sigma^K_{k=1} \beta_k y_k/\lambda_k)\exp(\Sigma^K_{k=1}
\beta_k^2\tau_k^2/2\lambda_k^2)
\end{align}
where the last step uses the fact that the expectation of the exp-sum of error terms is equivalent to the product of expectations of independent log-normal random variables.

With Eqs. \ref{eq:latentvariance2} and \ref{eq:latentvariance4}, we can obtain an expression that is analogous to Eq. \ref{eq:latentratio}, but fully expressed in terms of observable measurements (common factors are omitted):
\begin{align}
\frac{
    \text{Var} \left[ y_0\mid (y_1+1),y_{\{2,\dots,K\}} \right] - \tau_0^2
}{
    \text{Var} \left[ y_0\mid y_1,y_{\{2,\dots,K\}} \right] - \tau_0^2
} &= \frac{
    \text{Var} \left[ \zeta_0\mid (y_1+1),y_{\{2,\dots,K\}} \right]
}{
    \text{Var} \left[ \zeta_0\mid y_1,y_{\{2,\dots,K\}} \right]
} \nonumber\\ 
&= \frac{
    \exp(\beta_1 (y_1 + 1)/\lambda_1 + \Sigma^K_{k=2} \beta_k y_k/\lambda_k)
}{
    \exp(\beta_1 y_1/\lambda_1 + \Sigma^K_{k=2} \beta_k y_k/\lambda_k)
} \nonumber\\
&= \exp(\beta_1/\lambda_1).
\end{align}
Hence, observable changes in the variance of the measurement $y_0$ across measurements $y_k$ of the latent variance predictors provide sufficient information to estimate the coefficients and no additional identification constraints are required because $\tau^2_0$ and $\lambda_1$ are already identified by the usual constraints on the measurement model.

\subsection{Estimation}
\label{sec:estimation}

We use Bayesian inference for model fitting. At a high level, the process consists of first specifying a \emph{prior distribution} (further discussed in the next subsection), which describes our state of knowledge before seeing the data, and combining it with the data-informed model likelihood to obtain a \emph{posterior distribution}, which represents our updated state of knowledge about the parameters' values. An accessible introduction to Bayesian inference can be found in \textcite{johnsonBayesRulesIntroduction2022}.

Calculating the posterior distribution is the main challenge during inference as the expression involves a high-dimensional integral which will not have a closed-form beyond a few special cases; hence, it becomes necessary to resort to numerical methods. We use Markov chain Monte Carlo (MCMC), specifically adaptive Hamiltonian Monte Carlo as implemented in the Stan probabilistic programming language \parencite{hoffmanNoUTurnSamplerAdaptively2014,standevelopmentteamStanModelingLanguage2023}. Adaptive HMC is a class of efficient algorithms that can accurately sample complicated parameter spaces and Stan is a well-tested project that is freely available for all major operating systems. All MCMC algorithms produce sequences of samples (known as \emph{chains}) from the target distribution as its output, which we can then directly use to obtain estimates of parameter means, credibility intervals, transformations, and other quantities of interest \parencite{gelmanBayesianDataAnalysis2014}.

\subsection{Prior specification}

In the ideal Bayesian workflow, all model parameters are given priors that represent some state of knowledge which will be updated through the likelihood as new data arrives. The purpose of this paper, however, is to investigate the set of conditions under which our approach can produce useful results. Hence, we adopt the \emph{minimalist} position \parencite{gelmanPriorCanGenerally2017} for all simulations, i.e. we attempt to identify the weakest priors for each model that will still produce reliable inferences. The criteria we use to assess reliability are described in Section \ref{sec:diagnostics} and the specific priors are introduced along with their corresponding models in Section \ref{sec:results}. Readers looking for practical advice on how to set priors for SEM can find an excellent resource in \textcite{vanerpBayesianStructuralEquation2020} and \textcite{winterIllustratingValuePrior2023}.

\section{Simulations}
\label{sec:simstudy}

We investigated the viability of our approach through statistical simulation. Specifically, we tested four structural models that are likely to be relevant for practitioners: a simple two-factor model, a model with mediators, a model with interactions and a model with a sequential structure. The metrics used for assessment are introduced next, followed by a description of the computational setup, and then each model is presented together with the respective results.

\subsection{Model diagnostics}
\label{sec:diagnostics}
\subsubsection{Convergence}

While MCMC methods can work well in practice, convergence to a target distribution is an asymptotic property, so it is always necessary to verify convergence empirically \parencite{burknerModelsAreUseful2023}. This can be achieved by running multiple chains with randomized initial values and then examining whether they exhibit similar distributions; one commonly recommended convergence diagnostic is the \emph{potential scale reduction factor} $\hat R$ (often just called "Rhat"). Briefly, it compares the variance between and within chains as a proxy for convergence and returns a value in $[1,\infty)$, where values closer to 1 indicate the chains have more similar distributions. A detailed treatment can be found in \textcite{vehtariRankNormalizationFoldingLocalization2021}, where they also provide the recommendation to consider $\hat R \le 1.01$ as a reliable indicator of convergence. For the purpose of our simulation study, we relaxed the threshold to 1.05, as we have access to the ground truth values and therefore were able to verify that posterior estimates retained acceptable quality up to that point.

\subsubsection{Calibration}

\begin{figure}[hbt]
\includegraphics[width=\textwidth]{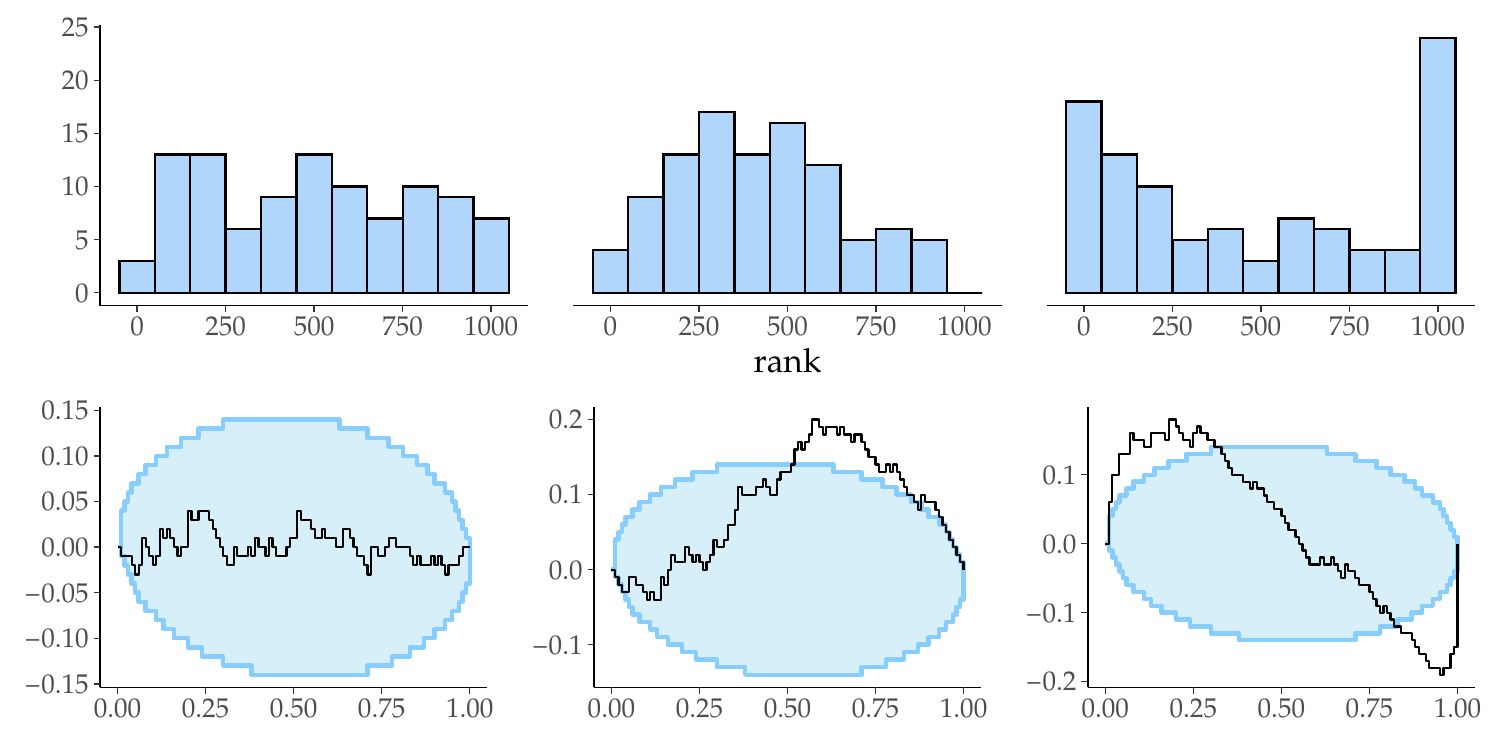}
\caption{Simulation-based rank histograms (top) and corresponding empirical cumulative distribution function (ECDF) difference plots (bottom) for three hypothetical quantities of interest. The blue areas in the ECDF difference plots indicate 95\%-confidence bands under the assumptions of uniformity and thus allow for a null-hypothesis significance test of self-consistent calibration. Left: A well-calibrated quantity. Center: A miscalibrated quantity with too many lower ranks indicating a positive bias in the estimated posteriors. Right: A miscalibrated quantity with too many extreme ranks indicating overconfident posteriors (i.e., variance underestimated).}
\label{fig:sbc}
\end{figure}

Convergence alone does not tell us whether our MCMC draws provide a good approximation to the true posterior distribution. However, we can use the draws themselves to diagnose the quality of our approximation if we also have knowledge of the true data-generating distribution; this is the key idea behind \emph{Simulation-Based Calibration} (SBC; \cite{taltsValidatingBayesianInference2020}). For this method, one samples parameters from the prior which are passed to the likelihood for data generation, the model is then fit over the resulting datasets, and the sum of ranks of the posterior draws relative to the true value is calculated; when a uniform distribution of rank sums is recovered, our posterior is said to be \emph{calibrated} (explained below). To assess uniformity, we used the graphical tests proposed in \textcite{sailynojaGraphicalTestDiscrete2022} (see Figure~\ref{fig:sbc}).

Calibration in the context of SBC is, strictly speaking, a statement about the expected coverage of posterior intervals over the joint distribution of data and parameters. In practice, this means that it can readily detect posterior approximations that consistently under-/overestimate the location or uncertainty that the true posterior would output for a given parameter; however, it can miss less obvious mismatches and hence does not provide a global guarantee of correctness \parencite{modrakSimulationBasedCalibrationChecking2023}. Fortunately, the procedure can be augmented with data-dependent quantities to provide a more stringent test; in particular, we also test the model likelihood, which greatly increases the sensitivity of the test as also demonstrated by \citeauthor{modrakSimulationBasedCalibrationChecking2023}.

\subsubsection{Effective sample size and efficiency}

Even if the model is calibrated and has converged, we only have a finite sample of MCMC draws from the posterior, so we must ensure that the estimation error is small enough to give us reliable inference. It is also necessary to account for the autocorrelation that is often present in the chains as this further reduces their information content; Effective Sample Size (ESS) is a diagnostic that addresses this by estimating the number of independent draws that the information in our chains is equivalent to (Section 11.5, \cite{gelmanBayesianDataAnalysis2014}). We consider an ESS of at least 100 per independent MCMC chain to be sufficient for reliable estimation and separately report bulk ESS and tail ESS, as suggested by \textcite{vehtariRankNormalizationFoldingLocalization2021}.

As having a high enough ESS is a prerequisite for accurate inference, a question of practical importance is how long one has to run a model for in order to achieve the desired precision. We calculate ESS per second (ESS/s) as it provides a simple measure of sampling efficiency for each model. However, this will vary considerably depending on the priors used, the data at hand, and the computer one uses to fit the model; the intent here is only to determine whether the models can run in a reasonable time.

\subsubsection{Parameter recovery}

To evaluate parameter recovery, we use bias and the Root Mean Squared Error (RMSE). Given a set $\theta^{(s)}$ of $S$ posterior draws and a true value $\theta^{*}$, we have
\begin{align}
\text{Bias} = \frac{1}{S}\sum_{s=1}^S \theta^{(s)} - \theta^{*},\\
\text{RMSE} = \sqrt{\frac{1}{S}\sum_{s=1}^S (\theta^{(s)} - \theta^{*})^2}.
\label{eq:rmse}
\end{align}
Posterior means will almost surely (in the formal sense) have non-zero bias whenever proper priors are used. However, they are also consistent estimators and we show that, for our models, this leads to the bias being negligible. As bias does not account for posterior uncertainty, we also report RMSE because it provides an overall indication of estimation error by incorporating both posterior bias and variance into a single measure. This relation can be shown explicitly by rearranging Eq.~\ref{eq:rmse}:
\begin{align}
\text{RMSE}^2 =  \underbrace{\frac{1}{S}\sum_{s=1}^S(\theta^{(s)} - \bar\theta)^2}_{\text{Variance}} + \underbrace{(\bar\theta - \theta^{*})^2}_{\text{Bias}^2},
\end{align}
where $\bar\theta = (1/S)\Sigma_{s=1}^S\theta^{(s)}$ is the posterior mean.

\subsection{Computational setup}
\label{sec:compsetup}

The workflow of our simulation study can be summarized in four key steps:

\begin{enumerate}
\item For each model, we found a relatively tight \emph{generative} prior distribution, such that parameter vectors drawn from the joint prior could be used to simulate datasets without degeneracies (e.g., without values close to zero for variances or factor loadings) with high probability.
\item We drew 250 parameter vectors from the generative prior and with each of these, we subsequently generated a dataset of 500 observations from the model likelihood.
\item We fitted the model twice, using two different priors, for each of the generated datasets: first with the generative prior itself and second with a much wider, weakly informative prior.
\item Finally, we processed the resulting posterior samples to obtain model diagnostics in the following manner:
\begin{itemize}
    \item Convergence was examined in both sets of models. Below, we only report convergence for the models with the weakly informative prior, as convergence for the models with the generative prior was always superior (see \hyperref[sec:appendix1]{Appendix A}).
    \item Calibration was assessed on the models that used the generative prior, because only there SBC is valid.
    \item Metrics for parameter recovery and sampling efficiency were calculated from the models with the weakly informative prior. We excluded models with any $\hat R > 1.05$ from this calculation to avoid artefacts caused by clear non-convergence.
\end{itemize}
\end{enumerate}

Our simulations were fully implemented using the \texttt{R} programming language \parencite{rcoreteamLanguageEnvironmentStatistical2023}. We specified our models in the \texttt{brms} package \parencite{burknerBrmsPackageBayesian2017}, which provides a user-friendly interface for generation of Stan code, and wraps the \texttt{cmdstanr} and \texttt{posterior} packages, which respectively provide functions for interfacing with Stan itself and for extracting model diagnostics \parencite{gabryCmdstanrInterfaceCmdStan2023,burknerPosteriorToolsWorking2023}. Functions needed for dataset generation, linking true parameter values to specific fits and plotting calibration diagnostics were provided by the
\texttt{SBC} package \parencite{kimSBCSimulationBased2023}. To facilitate reproducibility, the simulation pipeline itself was built to run via the \texttt{targets} package \parencite{landauTargetsPackageDynamic2021}. The full code is available at an online repository.\footnote{See the folder \texttt{simulation-study} at \url{https://github.com/bdlvm-project/gdsem-paper}}
The complete pipeline was run on a MacBook Pro with M2 chip, where it took approximately 42 hours to complete.

\subsection{Results}
\label{sec:results}

To aid visualisation of the models that follow, we introduce a novel graphical representation to complement the previously established notation. As our models allow the mean ($\mu$) and standard deviation ($\sigma$) of a latent variable distribution to be independently influenced by other latent variables, we extend the usual path diagram notation to show both of these parameters explicitly.

The diagram for a single latent variable with a five-item measurement model is shown in Figure~\ref{fig:model0}. This corresponds to the following statistical model:
\begin{align}
\begin{split}
\zeta_1 &\sim \text{Normal}(\mu_1, \sigma_1)\\
y_{1mi} &\sim \text{Normal}(\lambda_{1m} \, \zeta_{1i}, \tau_{1m}),
\end{split}
\end{align}
with ${m \in \{1,\dots,5\}}$ and ${\lambda_{11} = 1}$ for identification.
All the models in this section use five items per latent variable and a unit factor loading identification constraint, as in the example above. For brevity, we omit the measurement models in the descriptions that follow.

\begin{figure}[H]
		\centering
		\begin{tikzpicture}[scale=0.9]
		% nodes
		\node[draw,circle, minimum size=45pt] (zeta1) at (0,0) {};
            \node[draw,circle, dotted] (mu) at (-0.4,0.2) {$\mu$};
            \node[draw,circle, dotted] (sigma) at (0.4,0.2) {$\sigma$};
            \node (label) at (0.0, -0.4) {$\zeta_1$};
            % items zeta_1
		\node[draw] (y1) at (-2,-3) {$y_{1_1}$};
            \node[draw] (y2) at (-1,-3) {$y_{1_2}$};
		\node[draw] (y3) at (0,-3) {$y_{1_3}$};
		\node[draw] (y4) at (1,-3) {$y_{1_4}$};
            \node[draw] (y5) at (2,-3) {$y_{1_5}$};
            % connect items and node1
		\draw [->, >=latex] (zeta1) to node [midway,left] {1} (y1.north);%--(y1.north);
		\draw [->, >=latex] (zeta1)--(y2.north);
		\draw [->, >=latex] (zeta1)--(y3.north);
            \draw [->, >=latex] (zeta1)--(y4.north);
            \draw [->, >=latex] (zeta1)--(y5.north);
            % Residuals
            \draw[<->, >=latex] (y1) to [out=250,in=290,looseness=7] (y1);
		\draw[<->, >=latex] (y2) to [out=250,in=290,looseness=7] (y2);
		\draw[<->, >=latex] (y3) to [out=250,in=290,looseness=7] (y3);
		\draw[<->, >=latex] (y4) to [out=250,in=290,looseness=7] (y4);
		\draw[<->, >=latex] (y5) to [out=250,in=290,looseness=7] (y5);
            			
\end{tikzpicture}
		\caption{Extended path diagram notation which explicitly shows the parameters ($\mu$ and $\sigma$) that determine the latent variable's distribution.}
		\label{fig:model0}
\end{figure}

\subsubsection{Two-factor model}

We start with a two-factor model as this is the simplest setup where we can have a latent variance that is conditional on the value of another latent variable. The mathematical notation for the model is
\begin{align}
\begin{split}
\zeta_1 &\sim \text{Normal}(0, \sigma_1)\\
\zeta_2 &\sim \text{Normal}(\mu_2, \sigma_2)\\
\mu_2 &= \beta_{1 \mu_2} \zeta_1\\
\log \sigma_2 &=
\beta_{0 \sigma_2} + \beta_{1 \sigma_2} \zeta_1.
\end{split}
\end{align}
The path diagram representation is shown in Figure~\ref{fig:model1}. The priors used for generation and fitting are described in Table~\ref{tab:sim01prior}.

\begin{figure}[H]
		\centering
		\begin{tikzpicture}[scale=0.9]
		% nodes
		\node[draw,circle, minimum size=45pt] (zeta1) at (0,0) {};
            \node[draw,circle, dotted] (mu) at (-0.4,0.2) {$\mu$};
            \node[draw,circle, dotted] (sigma) at (0.4,0.2) {$\sigma$};
            \node (label) at (0.0, -0.4) {$\zeta_1$};
		\node[draw,circle, minimum size=45pt] (zeta2) at (3,0) {};
            \node[draw,circle, dotted] (mu2) at (2.6,0.2) {$\mu$};
            \coordinate (mupoint) at (2.5,0.3) {};
            \node[draw,circle, dotted] (sigma2) at (3.4,0.2) {$\sigma$};
            \coordinate (sigmapoint) at (3.32, 0.3);
            \node (label) at (3.0, -0.4) {$\zeta_2$};
            % connect nodes
		\draw [->, bend left=25, >=latex] (zeta1) to node [midway,above] {} (mupoint);
            \draw [->, bend left=35, >=latex] (zeta1) to node [midway,above] {} (sigmapoint);
			
\end{tikzpicture}
		\caption{Two-factor model. $\zeta_1$ influences both the mean and standard deviation of $\zeta_2$.}
		\label{fig:model1}
\end{figure}

\begin{table}
\centering
    \begin{tblr}{l c c c}
    \hline
    Parameter type & Notation & Generative prior & Weakly informative prior \\
    \hline
    Latent mean & &\\
    \quad Slope & $\beta_{1\mu_2}$ & \text{Normal}(1,\,0.3) & \text{Normal}(0,\,2.5)\\
    \hline[dashed]
    Latent std. dev.& &\\
    \quad Initial & $\sigma_1$ & $\text{Gamma}_{[0.7,\infty)}(11,11)$ & \text{Gamma}(5,\,5) \\
    \quad Intercept & $\beta_{0\sigma_2}$ & $\text{Exp-Gamma}(11,11)$ & \text{Exp-Gamma}(5,\,5) \\
    \quad Slope & $\beta_{1\sigma_2}$ & \text{Normal}(0.15,\,0.05) & \text{Normal}(0,\,0.5) \\
    \hline[dashed]
    Item parameters& &\\
    \quad Factor loadings & $\lambda$ & \text{Normal}(1,\,0.3) & \text{Normal}(0,\,2.5)\\
    \quad Error std. dev. & $\tau$ & $\text{Normal}_{[0.3,\infty)}(0.5,\,0.15)$ & Gamma(2.5,\,5)\\
    \hline
    \end{tblr}
\caption{Prior specifications for the two-factor model.}
\label{tab:sim01prior}
\end{table}

\begin{figure}[H]
\includegraphics[width=\textwidth]{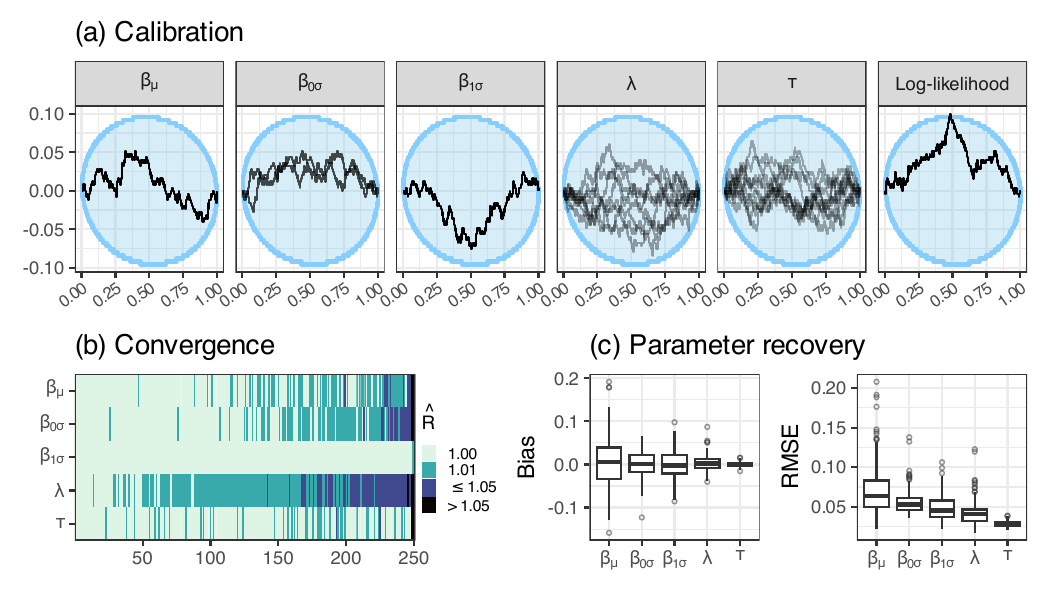}
\caption{Simulation diagnostics for the two-factor model. (a) ECDF difference plots. Curves are overlaid when there are multiple parameters of the same type. (b) Heatmap showing the average $\hat R$ for each parameter type in each simulation. Simulations are arranged in ascending order across the x-axis according to their overall mean $\hat R$. (c) Box plots of the error distribution for average bias and average RMSE per simulation and parameter type. Simulations with convergence issues (any parameter with $\hat R > 1.05$) were excluded.}
\label{fig:dx01}
\end{figure}

The calibration plots in Figure~\ref{fig:dx01}a show that all model parameters and the log-likelihood are well-calibrated.

The heatmap in Figure~\ref{fig:dx01}b shows that about two-thirds of the simulations produced estimates with $\hat R$ between 1.00 and 1.01. Factor loadings ($\lambda$) appear to be the parameters most likely to present suboptimal convergence but major problems are rare, with only two simulations having $\hat R >$ 1.05. We did not find any feature of the simulations that distinctly explained the variation in $\hat R$ which suggests it is primarily caused by variations in the random initializations of parameter values. Such occasional convergence issues are known to occur when fitting Bayesian SEMs in general so attempting to refit with a different seed is an advisable first step.\footnote{For instance, this is also recommended in the \href{https://ecmerkle.github.io/blavaan/articles/summaries.html\#convergence}{documentation} for Bayesian SEM package \texttt{blavaan}.}.

Finally, our parameter recovery plots (Figure~\ref{fig:dx01}c) show that bias is negligible across all model parameters. The values for RMSE need to be interpreted in the context of each parameter's relevant scale: for the slope on the standard deviation $\beta_{1\sigma}$, the parameter we are most interested in, an RMSE of 0.05 is low enough to expect that the model can be usefully employed to obtain directional estimates. This result is encouraging given that it comes from datasets with only 500 observations; studies with a larger sample size would be able to produce even more precise inferences.

\subsubsection{Mediation model}

One common use of SEM is mediation analysis, which allows researchers to quantify direct and indirect effects for a given variable. We show that our approach allows investigating mediation for effects on both means and standard deviations by fitting the following model:
\begin{align}
\begin{split}
\zeta_1 &\sim \text{Normal}(0, \sigma_1)\\
\zeta_2 &\sim \text{Normal}(\mu_2, \sigma_2)\\
\mu_2 &= \beta_{1 \mu_2} \zeta_1\\
\log \sigma_2 &=
\beta_{0 \sigma_2} + \beta_{1 \sigma_2} \zeta_1\\
\zeta_3 &\sim \text{Normal}(\mu_3, \sigma_3)\\
\mu_3 &= \beta_{1 \mu_3} \zeta_1\\
\log \sigma_3 &=
\beta_{0 \sigma_3} + \beta_{1 \sigma_3} \zeta_1\\
\zeta_4 &\sim \text{Normal}(\mu_4, \sigma_4)\\
\mu_4 &= \beta_{1 \mu_4} \zeta_1 + \beta_{2 \mu_4} \zeta_2\\
\log \sigma_4 &=
\beta_{0 \sigma_4} + \beta_{1 \sigma_4} \zeta_1 + \beta_{2 \sigma_4} \zeta_3.
\end{split}
\end{align}
Here, $\zeta_1$ has a direct effect on the $\mu$ and $\sigma$ for $\zeta_2$, $\zeta_3$, and $\zeta_4$. Additionally, it has an indirect effect on $\mu$ of $\zeta_4$ through $\zeta_2$ and an indirect effect on $\sigma$ of $\zeta_4$ through $\zeta_3$. The path diagram is shown in Figure~\ref{fig:model2}. Priors for this model are shown in Table~\ref{tab:sim02prior}.

\begin{figure}[H]
		\centering
		\begin{tikzpicture}[scale=0.9]
		% nodes
            \node[draw,circle, minimum size=45pt] (zeta1) at (0,0) {};
            \node[draw,circle, dotted] (mu) at (-0.4,0.2) {$\mu$};
            \node[draw,circle, dotted] (sigma) at (0.4,0.2) {$\sigma$};
            \node (label) at (0.0, -0.4) {$\zeta_1$};
            
		\node[draw,circle, minimum size=45pt] (zeta2) at (5,0) {};
            \node[draw,circle, dotted] (mu2) at (4.6,0.2) {$\mu$};
            \coordinate (mu2point) at (4.5, 0.3);
            \coordinate (mu2point2) at (4.57, 0.05);
            \node[draw,circle, dotted] (sigma2) at (5.4,0.2) {$\sigma$};
            \coordinate (sigma2point) at (5.28, 0.23);
            \coordinate (sigma2point2) at (5.35, 0.33);
            \node (label) at (5.0, -0.4) {$\zeta_4$};

            \node[draw,circle, minimum size=45pt] (zeta2) at (2.5,-2) {};
            \node[draw,circle, dotted] (mu21) at (2.1,-1.8) {$\mu$};
            \coordinate (mu21point) at (2, -1.72);
            \node[draw,circle, dotted] (sigma21) at (2.9,-1.8) {$\sigma$};
            \coordinate (sigma21point) at (2.81, -1.72);
            \node (label) at (2.5, -2.4) {$\zeta_{2}$};

            \node[draw,circle, minimum size=45pt] (zeta3) at (2.5,2) {};
            \node[draw,circle, dotted] (mu22) at (2.1,2.2) {$\mu$};
            \coordinate (mu22point) at (1.95, 2.2);
            \node[draw,circle, dotted] (sigma22) at (2.9,2.2) {$\sigma$};
            \coordinate (sigma22point) at (2.76, 2.2);
            \node (label) at (2.5, 1.6) {$\zeta_{3}$};
            
            % connect nodes
		\draw [->, bend left=20, >=latex] (zeta1) to node [midway,above] {} (mu2point);
            \draw [->, bend left=30, >=latex] (zeta1) to node [midway,above] {} (sigma2point);
            \draw [->, bend right=10, >=latex] (zeta1) to node [midway,left] {} (mu21point);
            \draw [->, bend right=10, >=latex] (zeta1) to node [midway,left] {} (sigma21point);
            \draw [->, bend left=10, >=latex] (zeta1) to node [near end,above] {} (mu22point);
            \draw [->, bend left=10, >=latex] (zeta1) to node [near end,above] {} (sigma22point);
            \draw [->, bend left=10, >=latex] (zeta3) to node [midway,left] {} (sigma2point2);
            \draw [->, bend right=10, >=latex] (zeta2) to node [midway,above] {} (mu2point2);
	
\end{tikzpicture}
		\caption{Mediation model. $\zeta_1$ has direct and indirect effects on both of $\zeta_4$'s parameters.}
		\label{fig:model2}
\end{figure}
\begin{table}
\centering
    \begin{tblr}{l c c c}
    \hline
    Parameter type & Notation & Generative prior & Weakly informative prior \\
    \hline
    Latent mean & &\\
    \quad Slope & $\beta_{\geq 1\mu}$ & \text{Normal}(1,\,0.3) & \text{Normal}(0,\,2.5)\\
    \hline[dashed]
    Latent std. dev.& &\\
    \quad Initial & $\sigma_1$ & $\text{Gamma}_{[0.7,\infty)}(11,11)$ & \text{Gamma}(5,\,5) \\
    \quad Intercept & $\beta_{0\sigma_2},\,\beta_{0\sigma_3},\,\beta_{0\sigma_4}$ & $\text{Exp-Gamma}(11,11)$ & \text{Exp-Gamma}(5,\,5) \\
    \SetCell[r=2]{l} \quad Slope & $\beta_{1\sigma_2},\,\beta_{1\sigma_3},\,\beta_{2\sigma_4}$ & \text{Normal}(-0.15,\,0.05) & \SetCell[r=2]{c} \text{Normal}(0,\,0.5) \\
    \quad & $\beta_{1\sigma_4}$ & \text{Normal}(0.15,\,0.05) & \\
    \hline[dashed]
    Item parameters& &\\
    \quad Factor loadings & $\lambda$ & \text{Normal}(1,\,0.3) & \text{Normal}(0,\,2.5)\\
    \quad Error std. dev. & $\tau$ & $\text{Normal}_{[0.3,\infty)}(0.5,\,0.15)$ & Gamma(2.5,\,5)\\
    \hline
    \end{tblr}
\caption{Prior specifications for the mediation model.}
\label{tab:sim02prior}
\end{table}

\begin{figure}[H]
\includegraphics[width=\textwidth]{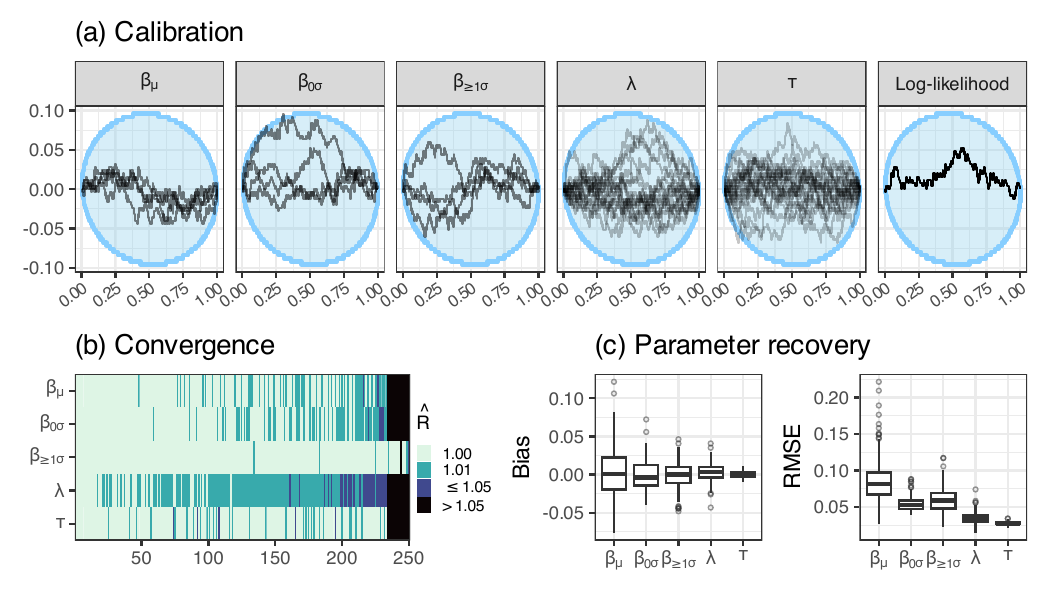}
\caption{Simulation diagnostics for the mediation model. (a) ECDF difference plots. Curves are overlaid when there are multiple parameters of the same type. (b) Heatmap showing the average $\hat R$ for each parameter type in each simulation. Simulations are arranged in ascending order across the x-axis according to their overall mean $\hat R$. (c) Box plots of the error distribution for average bias and average RMSE per simulation and parameter type. Simulations with convergence issues (any parameter with $\hat R > 1.05$) were excluded.}
\label{fig:dx02}
\end{figure}

The calibration plots in Figure~\ref{fig:dx02}a show good calibration across all test quantities. Convergence in Figure~\ref{fig:dx02}b again appears to be good in general; there are 17 simulations with $\hat R >$ 1.05, but we traced these cases back to generated datasets that produced very small variances in $\zeta_3$, which led to unstable estimates for factor loadings and error variances. We verified that refitting these cases with a different initialization was sufficient to resolve the issue (results not shown).

Results for parameter recovery in Figure~\ref{fig:dx02}c show no significant bias and RMSE is only slightly increased for all slopes (on both $\mu$ and $\sigma$) in the model compared to the previous two-factor model. This is an expected consequence of including mediators in the model, as having multiple paths for a given effect widens the range of coefficient values that are compatible with the data. In general, assessing mediation imposes increases in sample size and methodological complexity \parencite{rohrerThatLotProcess2022,montoyaSelectingBetweenSubjectDesign2023}.

\subsubsection{Interaction model}

We have mentioned in the Introduction that the ability to study moderation in general is a feature of interest for users of SEM frameworks. Our approach was created with the intent of providing more flexible models for moderation on latent variances, but we found that it can also accommodate moderation on structural paths with ease, since it can be represented through interactions between variables on the same linear predictor. We show this by fitting the moderation model depicted in Figure~\ref{fig:model3}, mathematically expressed as:
\begin{align}
\begin{split}
\zeta_1 &\sim \text{Normal}(0, \sigma_1)\\
\zeta_2 &\sim \text{Normal}(0, \sigma_2)\\
\zeta_3 &\sim \text{Normal}(0, \sigma_3)\\
\zeta_4 &\sim \text{Normal}(\mu_4, \sigma_4)\\
\mu_4 &= \beta_{1 \mu_4} \zeta_1 + \beta_{2 \mu_4} \zeta_1\zeta_2\\
\log \sigma_4 &=
\beta_{0 \sigma_4} + \beta_{1 \sigma_4} \zeta_1 + \beta_{2 \sigma_4} \zeta_1\zeta_3.
\end{split}
\end{align}

\begin{figure}[H]
\includegraphics[width=\textwidth]{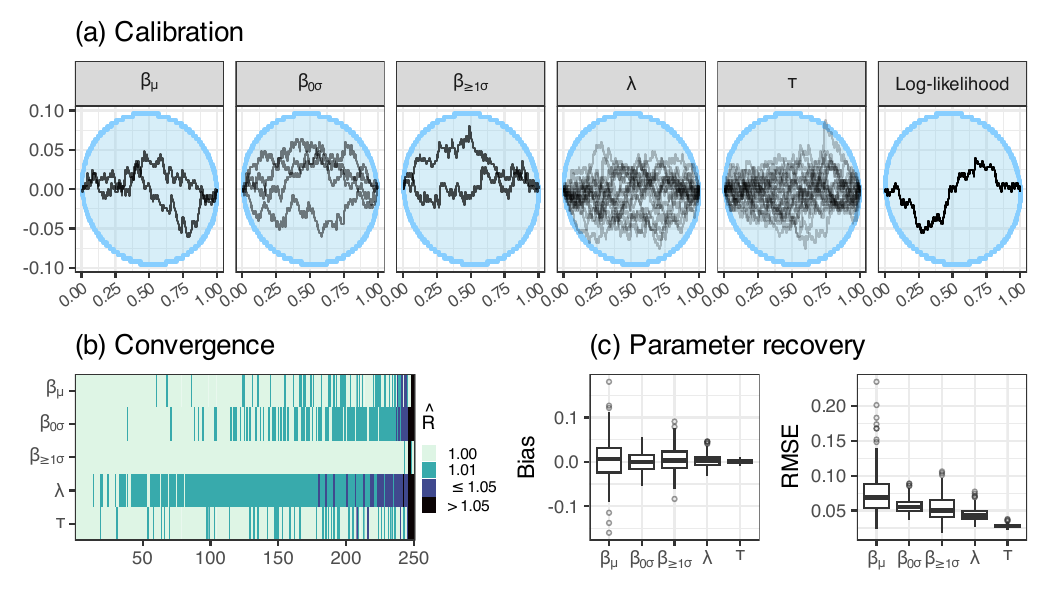}
\caption{Simulation diagnostics for the interaction model. (a) ECDF difference plots. Curves are overlaid when there are multiple parameters of the same type. (b) Heatmap showing the average $\hat R$ for each parameter type in each simulation. Simulations are arranged in ascending order across the x-axis according to their overall mean $\hat R$. (c) Box plots of the error distribution for average bias and average RMSE per simulation and parameter type. Simulations with convergence issues (any parameter with $\hat R > 1.05$) were excluded.}
\label{fig:dx03}
\end{figure}

The priors used to fit this model are given in Table~\ref{tab:sim03prior}.

\begin{figure}[H]
		\centering
		\begin{tikzpicture}[scale=0.9]
		% nodes
            \node[draw,circle, minimum size=45pt] (zeta1) at (0,0) {};
            \node[draw,circle, dotted] (mu) at (-0.4,0.2) {$\mu$};
            \node[draw,circle, dotted] (sigma) at (0.4,0.2) {$\sigma$};
            \node (label) at (0.0, -0.4) {$\zeta_1$};

            \node[draw,circle, minimum size=45pt] (zeta4) at (6,0) {};
            \node[draw,circle, dotted] (mu2) at (5.6,0.2) {$\mu$};
            \coordinate (mu2point) at (5.45, 0.2);
            \coordinate (mu2point2) at (5.47, 0.13);
            \node[draw,circle, dotted] (sigma2) at (6.4,0.2) {$\sigma$};
            \coordinate (sigma2point) at (6.3, 0.1);
            \coordinate (sigma2point2) at (6.35, 0.33);
            \node (label) at (6.0, -0.39) {$\zeta_4$};
  
            \node[draw,circle, minimum size=45pt] (zeta2) at (0,3.4) {};
            \node[draw,circle, dotted] (mu21) at (-0.4,3.6) {$\mu$};
            \node[draw,circle, dotted] (sigma21) at (0.4,3.6) {$\sigma$};
            \node (label) at (0, 3.0) {$\zeta_{2}$};
            
            \node[draw,circle, minimum size=45pt] (zeta3) at (6,3.4) {};
            \node[draw,circle, dotted] (mu22) at (5.6,3.6) {$\mu$};
            \node[draw,circle, dotted] (sigma22) at (6.4,3.6) {$\sigma$};
            \node (label) at (6, 3.0) {$\zeta_{3}$};

            \node[draw,circle, minimum size=32pt] (zeta22zeta1) at (4.3,1.7) {$\zeta_1 \zeta_{3}$};
            \node[draw,circle, minimum size=32pt] (zeta21zeta1) at (1.7,1.7) {$\zeta_1 \zeta_{2}$};
            
            % connect nodes
		\draw [->, bend right=25, >=latex] (zeta1) to node [midway,above] {} (mu2point2);
            \draw [->, bend right=25, >=latex] (zeta1) to node [midway,above] {} (sigma2point);
            \draw [->, bend right=25, >=latex, densely dashed] (zeta1) to node [midway,above] {} (zeta21zeta1);
            \draw [->, bend right=25, >=latex, densely dashed] (zeta2) to node [midway,above] {} (zeta21zeta1);
            \draw [->, bend right=25, >=latex, densely dashed] (zeta1) to node [midway,above] {} (zeta22zeta1);
            \draw [->, bend left=25, >=latex, densely dashed] (zeta3) to node [midway,above] {} (zeta22zeta1);
            \draw [->, bend left=15, >=latex] (zeta22zeta1) to node [midway,right] {} (sigma2point2);
            \draw [->, bend right=25, >=latex] (zeta21zeta1) to node [near end,below] {} (mu2point);
	
\end{tikzpicture}
		\caption{Interaction model. Dashed lines represent deterministic transformations; in this case, taking the product of two latent variables.}
		\label{fig:model3}
\end{figure}
\begin{table}
\centering
    \begin{tblr}{l c c c}
    \hline
    Parameter type & Notation & Generative prior & Weakly informative prior \\
    \hline
    Latent mean & &\\
    \SetCell[r=2]{l} \quad Slope & $\beta_{1\mu_4}$ & \text{Normal}(1,\,0.3) & \SetCell[r=2]{c} \text{Normal}(0,\,2.5)\\
    & $\beta_{2\mu_4}$ & \text{Normal}(0.5,\,0.3) & \\
    \hline[dashed]
    Latent std. dev.& &\\
    \quad Initial & $\sigma_1,\,\sigma_2,\,\sigma_3$ & $\text{Gamma}_{[0.7,\infty)}(11,11)$ & \text{Gamma}(5,\,5) \\
    \quad Intercept & $\beta_{0\sigma_4}$ & $\text{Exp-Gamma}(11,11)$ & \text{Exp-Gamma}(5,\,5) \\
    \SetCell[r=2]{l} \quad Slope & $\beta_{1\sigma_4}$ & \text{Normal}(0.1,\,0.05) & \SetCell[r=2]{c} \text{Normal}(0,\,0.5) \\
    \quad & $\beta_{2\sigma_4}$ & \text{Normal}(0.05,\,0.05) & \\
    \hline[dashed]
    Item parameters& &\\
    \quad Factor loadings & $\lambda$ & \text{Normal}(1,\,0.3) & \text{Normal}(0,\,2.5)\\
    \quad Error std. dev. & $\tau$ & $\text{Normal}_{[0.3,\infty)}(0.5,\,0.15)$ & Gamma(2.5,\,5)\\
    \hline
    \end{tblr}
\caption{Prior specifications for the interaction model.}
\label{tab:sim03prior}
\end{table}

The results in Figure~\ref{fig:dx03} show that the model is well-calibrated and has good convergence overall. Parameter recovery is in line with the previous models, showing no evidence of bias as well as low RMSE.

\subsubsection{Sequential model}

Model structures that contain longer sequences of latent variables can be relevant in studies that involve measurements over time \parencite{asparouhovDynamicStructuralEquation2018} or in those which seek to study detailed causal structures \parencite{zugnaAppliedCausalInference2022}. A thorough investigation of models in that space is well beyond the scope of this paper, but we considered it relevant to at least explore whether issues with our approach could become apparent only when fitting a longer sequence of dependencies between latent variables. The model we constructed additionally shows that transformations of latent variables are also supported in this approach (Figure~\ref{fig:model4}). Using mathematical notation:
\begin{align}
\begin{split}
\zeta_1 &\sim \text{Normal}(0, \sigma_1)\\
\zeta_j &\sim \text{Normal}(\mu_j, \sigma_j)\\
\mu_j &= \beta_{1 \mu_j} \zeta_{j-1}\\
\log \sigma_j &=
\beta_{0 \sigma_j} + \beta_{1 \sigma_j} \zeta^2_{j-1},
\end{split}
\end{align}
where $j \in \{2,3,4,5\}$ for this case. Model priors are shown in Table~\ref{tab:sim04prior}.

\begin{figure}[H]
		\centering
		\begin{tikzpicture}[scale=0.9]
		% nodes
            \node[draw,circle, minimum size=45pt] (zeta1) at (-8,0) {};
            \node[draw,circle, dotted] (mu) at (-8.4,0.2) {$\mu$};
            \node[draw,circle, dotted] (sigma) at (-7.6,0.2) {$\sigma$};
            \node (label) at (-8, -0.4) {$\zeta_1$};

            \node[draw,circle, minimum size=45pt] (zeta5) at (8,0) {};
            \node[draw,circle, dotted] (mu5) at (7.6,0.2) {$\mu$};
            \coordinate (mu5point) at (7.47, 0.1);
            \node[draw,circle, dotted] (sigma5) at (8.4,0.2) {$\sigma$};
            \coordinate (sigma5point) at (8.3, 0.3);
            \node (label) at (8, -0.4) {$\zeta_5$};

            \node[draw,circle, minimum size=45pt] (zeta2) at (-4,0) {};
            \node[draw,circle, dotted] (mu2) at (-4.4,0.2) {$\mu$};
            \coordinate (mu2point) at (-4.53, 0.1);
            \node[draw,circle, dotted] (sigma2) at (-3.6,0.2) {$\sigma$};
            \coordinate (sigma2point) at (-3.7, 0.3);
            \node (label) at (-4, -0.4) {$\zeta_2$};

            \node[draw,circle, minimum size=45pt] (zeta4) at (4,0) {};
            \node[draw,circle, dotted] (mu4) at (3.6,0.2) {$\mu$};
            \coordinate (mu4point) at (3.47, 0.1);
            \node[draw,circle, dotted] (sigma4) at (4.4,0.2) {$\sigma$};
            \coordinate (sigma4point) at (4.3, 0.3);
            \node (label) at (4, -0.4) {$\zeta_4$};

            \node[draw,circle, minimum size=45pt] (zeta3) at (0,0) {};
            \node[draw,circle, dotted] (mu3) at (-0.4,0.2) {$\mu$};
            \coordinate (mu3point) at (-0.53, 0.1);
            \node[draw,circle, dotted] (sigma3) at (0.4,0.2) {$\sigma$};
            \coordinate (sigma3point) at (0.3, 0.3);
            \node (label) at (0, -0.4) {$\zeta_3$};

            \node[draw,circle, minimum size=32pt] (zeta1^2) at (-6,0) {$\zeta_1^2$};
            \node[draw,circle, minimum size=32pt] (zeta2^2) at (-2,0) {$\zeta_2^2$};
            \node[draw,circle, minimum size=32pt] (zeta3^2) at (2,0) {$\zeta_3^2$};
            \node[draw,circle, minimum size=32pt] (zeta4^2) at (6,0) {$\zeta_4^2$};
  
            % connect nodes
		\draw [->, bend left=55, >=latex] (zeta1^2) to node [midway,right] {} (sigma2point);
            \draw [->, bend right=45, >=latex] (zeta1) to node [midway,right] {} (mu2point);
            \draw [->, bend left=55, >=latex] (zeta2^2) to node [midway,right] {} (sigma3point);
            \draw [->, bend right=45, >=latex] (zeta2) to node [midway,right] {} (mu3point);
            \draw [->, bend left=55, >=latex] (zeta3^2) to node [midway,right] {} (sigma4point);
            \draw [->, bend right=45, >=latex] (zeta3) to node [midway,right] {} (mu4point);
            \draw [->, bend left=55, >=latex] (zeta4^2) to node [midway,right] {} (sigma5point);
            \draw [->, bend right=45, >=latex] (zeta4) to node [midway,right] {} (mu5point);
            \draw [->, bend left=45, >=latex, densely dashed] (zeta1) to node [midway,above] {} (zeta1^2);
            \draw [->, bend left=45, >=latex, densely dashed] (zeta2) to node [midway,above] {} (zeta2^2);
            \draw [->, bend left=45, >=latex, densely dashed] (zeta3) to node [midway,above] {} (zeta3^2);
            \draw [->, bend left=45, >=latex, densely dashed] (zeta4) to node [midway,above] {} (zeta4^2);
			
\end{tikzpicture}
		\caption{Sequential model.}
		\label{fig:model4}
\end{figure}
\begin{table}
\centering
    \begin{tblr}{l c c c}
    \hline
    Parameter type & Notation & Generative prior & Weakly informative prior \\
    \hline
    Latent mean & &\\
    \quad Slope & $\beta_{1 \mu}$ & \text{Normal}(0,\,0.2) & \text{Normal}(0,\,2.5)\\
    \hline[dashed]
    Latent std. dev.& &\\
    \quad Intercept & $\beta_{0\sigma_1}$ & $\text{Gamma}_{[0.7,\infty)}(11,11)$ & \text{Gamma}(5,\,5) \\
    \quad Intercept & $\beta_{0\sigma_{\geq 2}}$ & $\text{Exp-Gamma}(11,11)$ & \text{Exp-Gamma}(5,\,5) \\
    \quad Slope & $\beta_{1 \sigma}$ & \text{Normal}(0,\,0.05) & \text{Normal}(0,\,0.5) \\
    \hline[dashed]
    Item parameters& &\\
    \quad Factor loadings & $\lambda$ & \text{Normal}(1,\,0.3) & \text{Normal}(0,\,2.5)\\
    \quad Error std. dev. & $\tau$ & $\text{Normal}_{[0.3,\infty)}(0.5,\,0.15)$ & Gamma(2.5,\,5)\\
    \hline
    \end{tblr}
\caption{Prior specifications for the sequential model.}
\label{tab:sim04prior}
\end{table}

Readers familiar with time series may note that setting an equality constraint on the coefficients would give the form of an autoregressive model, but testing this for lengths of time series representative of real applications would involve computational challenges that we do not aim to tackle here.

\begin{figure}[H]
\includegraphics[width=\textwidth]{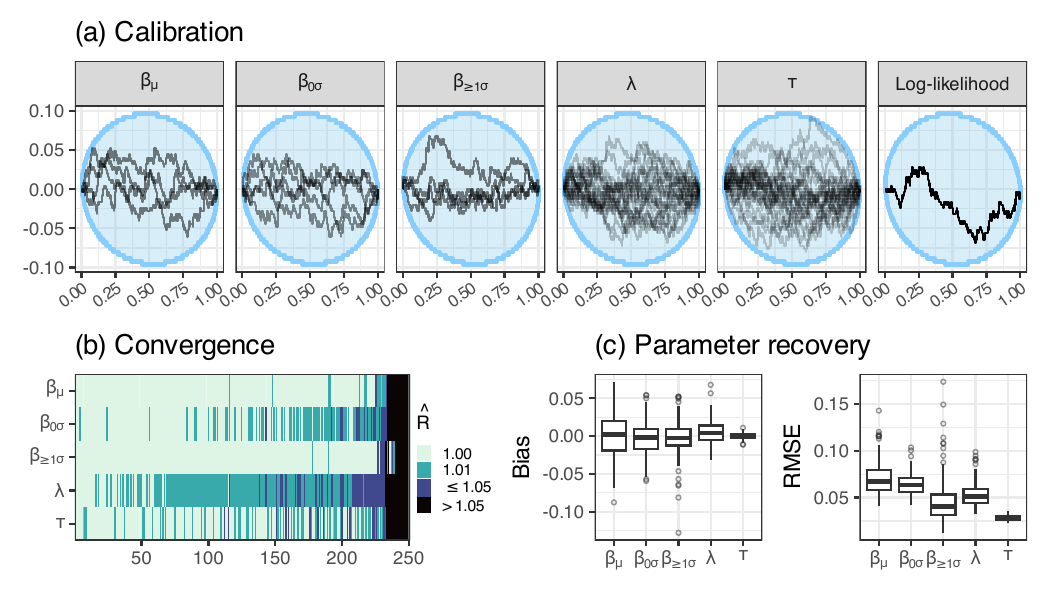}
\caption{Simulation diagnostics for the sequential model. (a) ECDF difference plots. Curves are overlaid when there are multiple parameters of the same type. (b) Heatmap showing the average $\hat R$ for each parameter type in each simulation. Simulations are arranged in ascending order across the x-axis according to their overall mean $\hat R$. (c) Box plots of the error distribution for average bias and average RMSE per simulation and parameter type. Simulations with convergence issues (any parameter with $\hat R > 1.05$) were excluded.}
\label{fig:dx04}
\end{figure}

The sequential structure of this model meant that, for certain parameter draws from the generative prior, there was a runaway increase in latent variance. Therefore, we had to drop one out of the 250 datasets due to overflowing variances which caused numerical errors. Similarly to the mediation model, there were also 18 datasets with unrealistic values for the variance parameters, which is reflected as a larger fraction of models with convergence issues in Figure~\ref{fig:dx04}. Nonetheless, calibration for this model was good, the vast majority of simulations converged well, and estimates don't show any major sign of bias. We did note a longer tail of high-RMSE estimates for the $\beta_{\geq 1\sigma}$ coefficients, but the bulk of simulations remained at levels comparable to the previously tested models.

\subsubsection{Comparing ESS}

\begin{figure}[H]
\includegraphics[width=\textwidth]{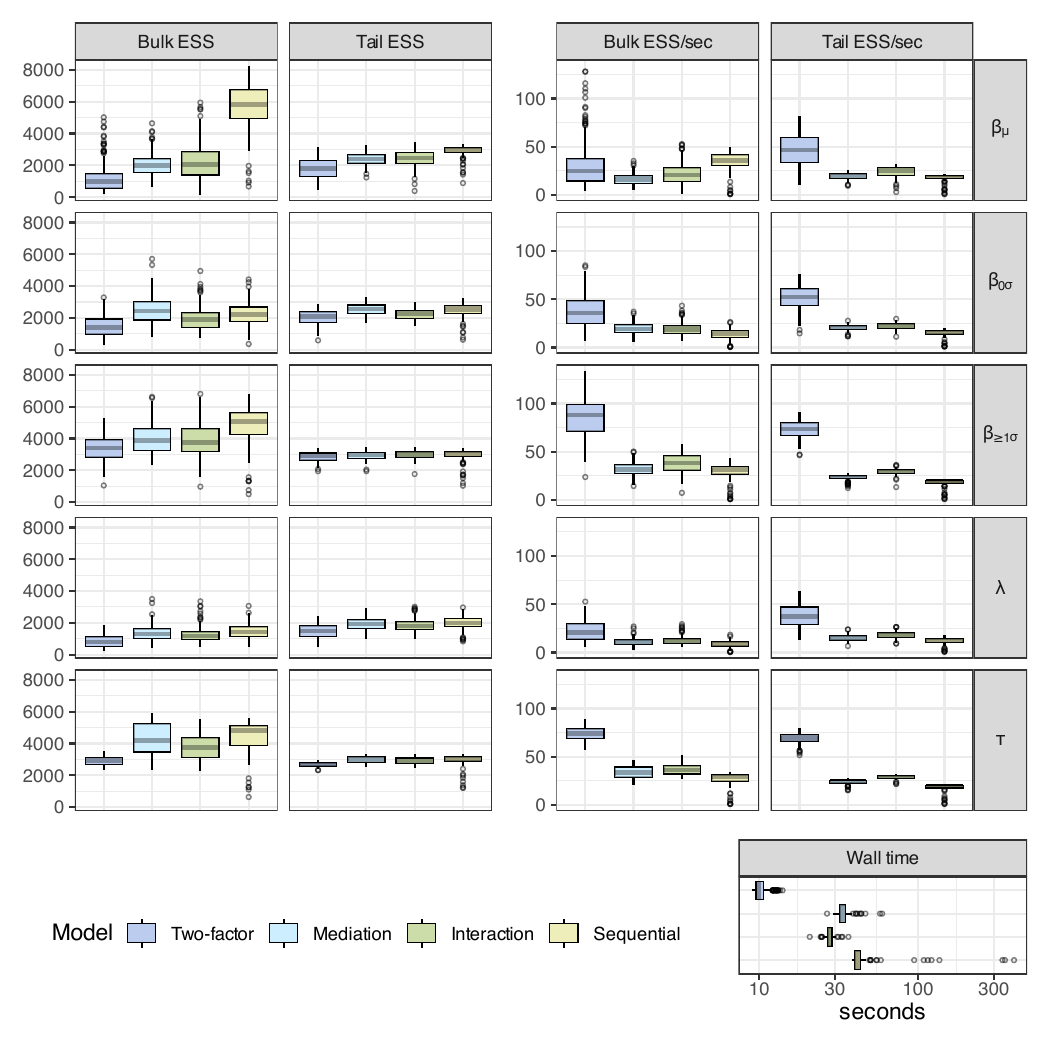}
\caption{Top two plots show the distribution of average effective sample size (ESS) and average ESS per second over all simulations, after excluding those with convergence issues (any parameter with $\hat R > 1.05$). Lower right corner is the distribution of wall time of those simulations.}
\label{fig:ess}
\end{figure}

All the diagnostics discussed so far are based on a finite amount of samples drawn from the posterior and so, in principle, are subject to estimation error. The left side of Figure~\ref{fig:ess} shows that across all models and parameters, the resulting ESS was well above the recommended threshold of 100 samples per chain, both for bulk and tail estimates, which establishes that our sample-based diagnostics are providing reliable information.

On the top right side of Figure~\ref{fig:ess} we depict the ESS per second, which is a  measure of the sampling efficiency of our models. As expected, increasing the number of latent variables in the model generally leads to decreases in sampling efficiency. Available alternatives for SEM based on marginal likelihoods do not allow latent moderation to be modeled with the same degree of flexibility as our approach, so a direct comparison cannot be made, but we consider that our conditional likelihood models are sufficiently fast for practical everyday use: the lower right corner of Figure~\ref{fig:ess} shows that the vast majority of models took less than a minute to fit.

\section{Case study}
\label{sec:casestudy}

We sought to address a substantively relevant research question using dataset from a real study in order to demonstrate the usefulness and flexibility of our model. We reached out to the authors of \textcite{maderEmotionalStabilityNeuroticism2023}, as their paper investigates the association between neuroticism and intra-person variation in negative affect, both of which can be conceptualized as latent variables.

A key finding of \citeauthor{maderEmotionalStabilityNeuroticism2023} is that floor effects must be accounted for to reliably detect the neuroticism-emotional variability association. They used a hierarchical distributional regression model, where each individual's mean negative affect score was treated as a censored outcome and both its mean and variance were regressed against their mean neuroticism score. This is already quite an advanced model, but it has some shortcomings as it ignores the uncertainty in the score means and assumes every item should be weighted equally. However, methods available at the time would not have allowed to properly model the latent nature of neuroticism and negative affect while simultaneously estimating how one affects the variability of the other.

One of the authors kindly provided us with access to a subset of the \emph{Goettingen Ovulatory Cycle Diaries 2} dataset.\footnote{Codebook available at \url{https://rubenarslan.github.io/gocd2/}} The recruitment and data collection process is described in \textcite{arslanRoutinelyRandomizePotential2021}. Briefly, women filled out a form upon recruitment, which included a personality questionnaire, and were subsequently invited to fill out an online diary every day for 70 days, which included items on loneliness, irritability, self-esteem, stress and mood. We used this set of items as measurements for the emotional affect construct and the personality items in the initial questionnaire for the neuroticism construct. The study used a planned missingness design, which allows us to drop all incomplete observations without risk of bias. This leaves a total of 1039 women, each observed between 1 and 11 times (mean: 2.5).

\begin{figure}[H]
\includegraphics[width=\textwidth]{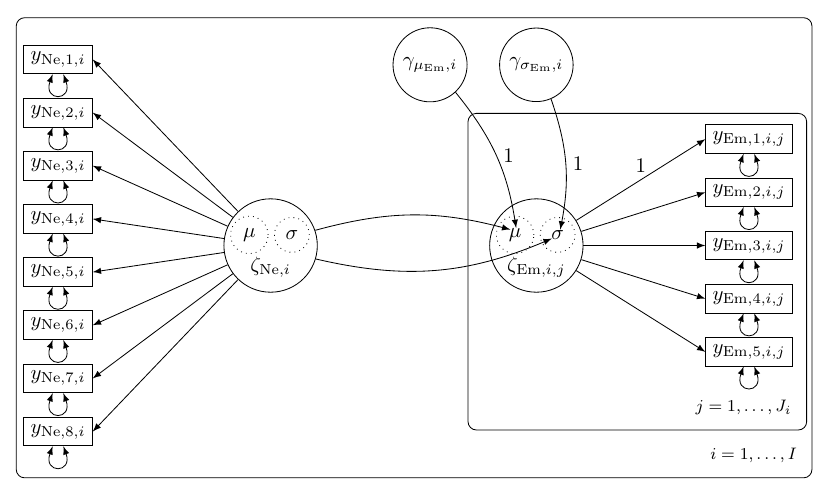}
		\caption{For subject $i$, the baseline measurement of neuroticism is used to predict the mean and variance of the emotional affect measurement taken at each time $j$.}
		\label{fig:casemodel}
\end{figure}

We aim to follow the model given in \textcite{maderEmotionalStabilityNeuroticism2023} as closely as possible. Therefore, we consider item responses at the extremes of the scale as censored and use a hierarchical structure to account for the repeated within-person observations. Abbreviating neuroticism as \emph{Ne} and emotional affect as \emph{Em}, our model can be notationally expressed as

\begin{align}
\begin{split}
\label{eq:casemodel}
\zeta_{\text{Ne},i} &\sim \text{Normal}(\mu_{\text{Ne}}, \sigma_{\text{Ne}})\\
\zeta_{\text{Em},i,j} &\sim \text{Normal}(\mu_{\text{Em},i}, \sigma_{\text{Em},i})\\
\mu_{\text{Em},i} &=
\beta_{0\mu_\text{Em}} + \beta_{1\mu_\text{Em}} \zeta_{\text{Ne},i} + \gamma_{\mu_\text{Em},i}\\
\log \sigma_{\text{Em},i} &=
\beta_{0\sigma_\text{Em}} + \beta_{1\sigma_\text{Em}} \zeta_{\text{Ne},i} + \gamma_{\sigma_\text{Em},i}\\
\gamma_{\mu_\text{Em},i} &\sim \text{Normal}(0, \sigma_{\mu_\text{Em}})\\
\gamma_{\sigma_\text{Em},i} &\sim \text{Normal}(0, \sigma_{\sigma_\text{Em}})\\
y^*_{\text{Ne},n,i} &\sim \text{Normal}(\nu_{\text{Ne},n}+\lambda_{\text{Ne},n} \, \zeta_{\text{Ne},i}, \tau_{\text{Ne},n})\\
y_{\text{Ne},n,i} &=
    \begin{cases}
    1 & y^*_{\text{Ne},n,i} \leq 1\\
    y^*_{\text{Ne},n,i} & 1 < y^*_{\text{Ne},n,i} < 5\\
    5 & 5 \leq y^*_{\text{Ne},n,i} \\
    \end{cases}
\\
y^*_{\text{Em},m,i,j} &\sim \text{Normal}(\nu_{\text{Em},m}+\lambda_{\text{Em},m} \, \zeta_{\text{Em},i,j}, \tau_{\text{Em},m})\\
y_{\text{Em},m,i,j} &=
    \begin{cases}
    0 & y^*_{\text{Em},m,i,j} \leq 0\\
    y^*_{\text{Em},m,i,j} & 0 < y^*_{\text{Em},m,i,j} < 4\\
    4 & 4 \leq y^*_{\text{Em},m,i,j} \\
    \end{cases}
\end{split}
\end{align}

where $i$ indexes individual participants, $j$ indexes their responses over the study duration, ${n \in \{1,\dots,8\}}$ indexes the items measuring neuroticism, ${m \in \{1,\dots,5\}}$ indexes the items measuring emotional affect, $\gamma$ represents person-specific random intercepts, and $y$ represents the observed (censored) responses to the questionnaires. The corresponding path diagram is shown in Figure~\ref{fig:casemodel}.

To obtain an identified model, we set ${\mu_{\text{Ne}} = \beta_{0\mu_\text{Em}} = 0}$ and ${\sigma_{\text{Ne}} = \lambda_{\text{Em},1} = 1}$. We chose the variance identification constraints pragmatically by fitting each latent variable in a separate model under each possible choice of constraint and picking the one which resulted in higher ESS for the remaining parameters, but in general one should also keep in mind the way constraints interact with priors \parencite{gravesNoteIdentificationConstraints2021}.

\begin{table}
\centering
    \begin{tblr}{l c c}
    \hline
    Parameter type & Notation & Prior \\
    \hline
    Latent mean & &\\
    \quad Slope & $\beta_{1\mu_\text{Em}}$ & \text{Normal}(0,\,2) \\
    \quad Std. dev. for varying intercept & $\sigma_{\mu_\text{Em}}$ & $\text{Half-normal}^+(0, 0.25)$ \\
    \hline[dashed]
    Latent std. dev.& &\\
    \quad Intercept & $\beta_{0\sigma_\text{Em}}$ & \text{Normal}(0,\,0.25) \\
    \quad Slope & $\beta_{1\sigma_\text{Em}}$ & \text{Normal}(0,\,0.25) \\
    \quad Std. dev. for varying intercept & $\sigma_{\sigma_\text{Em}}$ & $\text{Half-normal}^+(0, 0.25)$ \\
    \hline[dashed]
    Item parameters& &\\
    \quad Intercepts (centered) & $\nu$ & Student-t(3,\,0,\,2.5) \\
    \quad Factor loadings & $\lambda$ & \text{Normal}(0,\,2) \\
    \quad Error std. dev. & $\tau$ & Gamma(5,\,5) \\
    \hline
    \end{tblr}
\caption{Prior specification for the case study.}
\label{tab:caseprior}
\end{table}

We fit the model using weakly informative priors for all parameters as described in table \ref{tab:caseprior}. It was implemented using the \texttt{brms} package and code is available online.\footnote{See the folder \texttt{case-study} at \url{https://github.com/bdlvm-project/gdsem-paper}} Posterior means and credible intervals are shown in Figure~\ref{fig:casepost}. The key parameter of interest $\beta_\sigma$ was well-estimated with an Rhat of 1.00 and ESS above 1200 for bulk and tail, and had a posterior mean of 0.11 with a [0.05, 0.18] 95\% credible interval; this is consistent with \textcite{maderEmotionalStabilityNeuroticism2023}, which pooled 13 studies to produce an estimate of 0.10 [0.07, 0.13]. Strictly speaking, the parameters cannot be directly compared as the free factor loadings in our model lead to items being weighed differently, but investigating measurement invariance is beyond the scope of this example (however, see \cite{robitzschWhyFullPartial2023}).

\input{fig_casestudy_posteriors}

This case study illustrates how our approach opens the door for more truthful modeling of measurement processes. The results show that the latent parameters can be well-estimated even as part of a more complex model structure. Furthermore, the implementation cost of adding censoring and varying intercepts for this analysis was essentially negligible as these features were already available in the \texttt{brms} package. This extensibility arises from the underlying conditional likelihood formulation of the latent variable model and it should offer more flexibility for researchers who wish to build sophisticated models without the need to develop distinct, SEM-specific implementations.

\section{Discussion}
\label{sec:discussion}

In this article, we developed a Gaussian distributional SEM framework for flexible estimation of latent variable models that include latent moderators of both latent means and latent variances. We achieved this by using a Bayesian framework, which confirms the suggestion put forward by \textcite{dekortStudyingStrengthPrediction2017} that Bayesian estimation should be a viable approach for handling latent heteroscedasticity within more complex model structures.

Our simulation study used four distinct model structures to test the reliability of the estimates obtained through the conditional likelihood approach. Although our results show that all model parameters were well-calibrated, we wish to emphasize that the SBC procedure only provides information for the parameter region covered by the generative prior, and the ones we used were relatively narrow. We did not systematically investigate wider generative priors because they produced datasets with unrealistic properties and, as a result, lead to convergence issues too often to be reliably fitted. However, we did employ weakly informative priors for the assessment of convergence and parameter recovery with favorable results. Therefore, we anticipate our framework to function well across a broad range of prior specifications. That said, it is highly recommended that users employ prior predictive checks to ensure the appropriateness of their choices in any particular analysis (\cite{winterIllustratingValuePrior2023} provide a practical illustration of this technique in the specific context of SEM).

One important practical consideration that we did not address here is the impact of model misspecification on the resulting inferences. Results from \textcite{dekortStudyingStrengthPrediction2017} show that biased estimates will be obtained in situations where heteroscedasticity and nonlinearity are simultaneously present. Given that the model examined by \citeauthor{dekortStudyingStrengthPrediction2017} can be seen as a special case of the structures we have considered here, we expect the same caveats to carry over.

For future research, the analogy with distributional regression directly suggests the possibility of using conditional likelihood SEMs to explore non-Gaussian distributions for latent variables, including  the specification of moderators on distributional parameters beyond the variance. As mentioned above, we also faced the challenge of specifying sensible generative priors while designing our simulation study. It has already been highlighted by \textcite{merkleOpaquePriorDistributions2023} that the default approach of using non-informative priors implies data-generating processes that are incompatible with the patterns that would motivate using SEM in the first place. However, during model validation (e.g., when performing SBC) one would also like to cover as much of the parameter space as possible. Therefore, a relevant direction may be to move away from the usual approach of specifying priors on each individual model parameter and instead explore methods that use information expressed on more intuitive scales to construct the implied prior on the parameter scale (e.g., \cite{aguilarIntuitiveJointPriors2023,bocktingSimulationBasedPriorKnowledge2023}). Another possibility that we recently became aware of is to keep non-informative priors while simultaneously introducing imaginary data in order to produce an updated prior (for an overview of the approach see \cite{ibrahimPowerPriorTheory2015}; we demonstrate an application to SEM in \cite{fazioGenerativeBayesianModeling2024}).

Finally, while we have shown that the performance of our Gaussian distributional SEMs is sufficient for practical everyday applications, there is certainly room for further optimization. For this paper we used the \texttt{brms}-generated Stan code as-is, but we are aware that applying a non-centered parametrization \parencite{papaspiliopoulosGeneralFrameworkParametrization2007} to the latent variables leads to noticeable performance gains, so it would be helpful to implement this as an option in \texttt{brms} itself. Alternatively, variational approximations can be used in place of MCMC for fast posterior estimation. Initial results for SEM estimation have been encouraging \parencite{dangFittingStructuralEquation2022} but the statistical performance of these approximate methods still needs to be studied in a wider range of scenarios. Another promising set of approaches are those from the field of simulation-based inference, in particular, machine learning-based methods of posterior estimation and amortized inference \parencite{cranmerFrontierSimulationbasedInference2020,radevBayesFlowLearningComplex2022,zammit-mangionNeuralMethodsAmortized2024}. These techniques offer much faster inference-time results at the upfront cost of an initial training phase, but we have not found works that show their specific application to SEM estimation at this time.

\section*{Acknowledgments}

This research was partially funded by Deutsche Forschungsgemeinschaft (DFG, German Research Foundation) Project 497785967. We thank Timo Stenz for producing the path diagrams shown in this paper and the anonymous reviewers who provided us with valuable feedback.

%--------------------------------------------------------%
%	REFERENCE LIST
%--------------------------------------------------------%

{\singlespacing
% BIBLATEX
\printbibliography}

% APPENDIX
\newpage
\section*{Appendix}
\pagenumbering{roman} 
% use letters instead of numbers for appendix
\renewcommand{\thesubsection}{\Alph{subsection}}
\renewcommand\thesection{\Roman{section}}
% change Figure/Table numbering for appendix
\renewcommand\thefigure{\thesubsection.\arabic{figure}} 
% restart counter
\setcounter{figure}{0}
\setcounter{table}{0}
\subsection{Additional diagnostics}\label{sec:appendix1}

\begin{figure}[H]
\includegraphics[width=\textwidth]{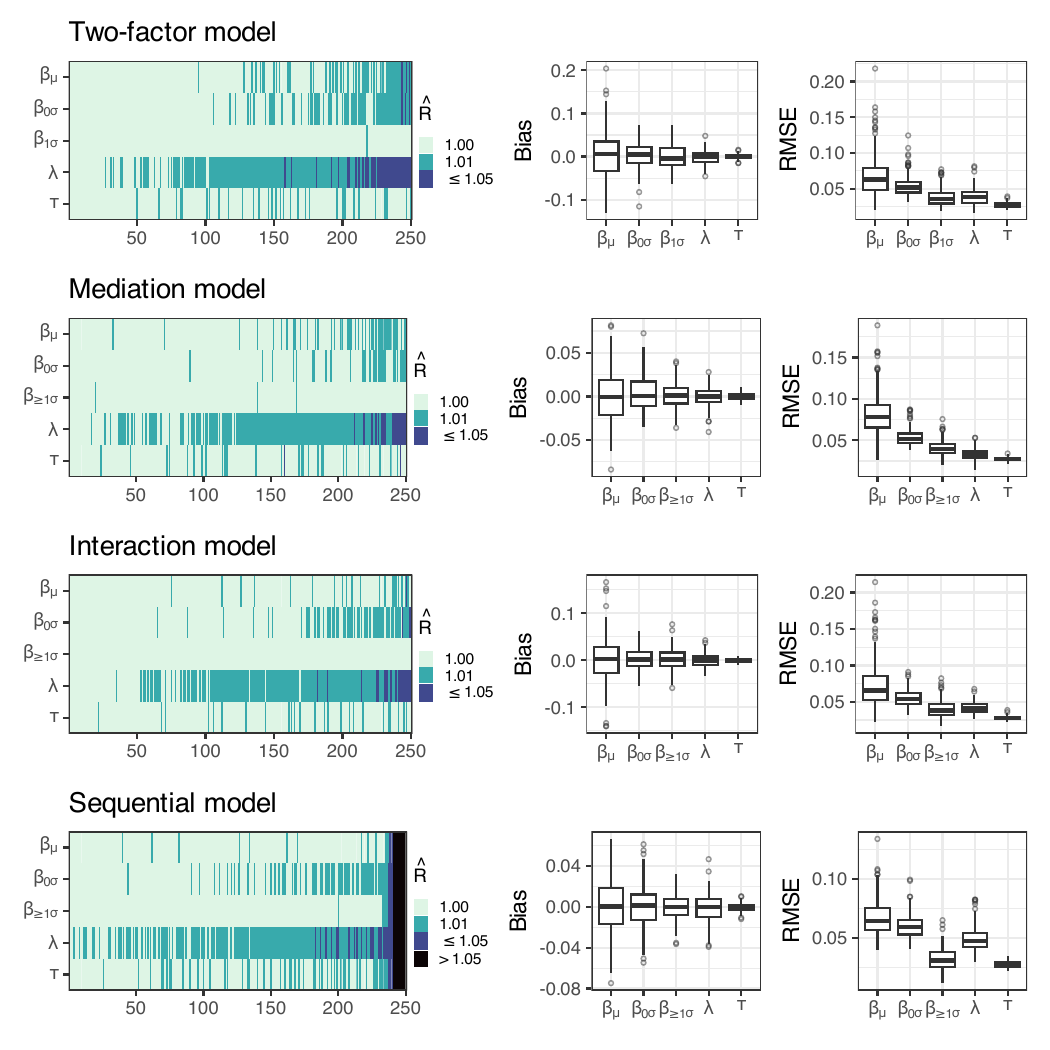}
\caption{Convergence and parameter recovery results for fits with the generative prior.}
\label{fig:appendix}
\end{figure}

%--------------------------------------------------------%
%	END DOCUMENT
%--------------------------------------------------------%

\end{document}